\documentclass[10pt, conference]{IEEEtran}
\IEEEoverridecommandlockouts
\usepackage{cite}
\usepackage{amsmath,amssymb,amsfonts}
\usepackage{algorithmic}
\usepackage{graphicx}
\usepackage{textcomp}
\usepackage{xcolor}
\def\BibTeX{{\rm B\kern-.05em{\sc i\kern-.025em b}\kern-.08em
    T\kern-.1667em\lower.7ex\hbox{E}\kern-.125emX}}

\usepackage{booktabs}
\usepackage{multirow}
\usepackage[T1]{fontenc}

\usepackage[most]{tcolorbox}
\tcbuselibrary{listings}

\begin{document}

\title{LLM-Powered Fully Automated Chaos Engineering: Towards Enabling Anyone to Build Resilient Software Systems at Low Cost}

% \author{
% \IEEEauthorblockN{Daisuke Kikuta}
% \IEEEauthorblockA{\textit{NTT, Inc., Japan} \\
% daisuke.kikuta@ntt.com}
% \and
% \IEEEauthorblockN{Hiroki Ikeuchi}
% \IEEEauthorblockA{\textit{NTT, Inc., Japan}\\
% hiroki.ikeuchi@ntt.com}
% \and
% \IEEEauthorblockN{Kengo Tajiri}
% \IEEEauthorblockA{\textit{NTT, Inc., Japan}\\
% kengo.tajiri@ntt.com}
% }
% \IEEEaftertitletext{\vspace{-1\baselineskip}}
\author{
\IEEEauthorblockN{Daisuke Kikuta, Hiroki Ikeuchi, Kengo Tajri}
\IEEEauthorblockA{\textit{NTT, Inc., Japan} \\
daisuke.kikuta@ntt.com}
}
\maketitle

\begin{abstract}
Chaos Engineering (CE) is an engineering technique aimed at improving the resilience of distributed systems.
It involves intentionally injecting faults into a system to test its resilience, uncover weaknesses, and address them before they cause failures in production.
Recent CE tools automate the execution of predefined CE experiments.
However, planning such experiments and improving the system based on the experimental results still remain manual.
These processes are labor-intensive and require multi-domain expertise.
% Proposal
To address these challenges and enable anyone to build resilient systems at low cost, this paper proposes ChaosEater, a system that automates the entire CE cycle with Large Language Models (LLMs).
It predefines an agentic workflow according to a systematic CE cycle and assigns subdivided processes within the workflow to LLMs.
ChaosEater targets CE for software systems built on Kubernetes.
Therefore, the LLMs in ChaosEater complete CE cycles through software engineering tasks, including requirement definition, code generation, testing, and debugging.
We evaluate ChaosEater through case studies on small- and large-scale Kubernetes systems.
The results demonstrate that it consistently completes reasonable CE cycles with significantly low time and monetary costs. 
Its cycles are also qualitatively validated by human engineers and LLMs.
% Our code can be found at the supplementary material folder.
\end{abstract}


\begin{figure*}[tb] \centering
    \includegraphics[width=\linewidth]{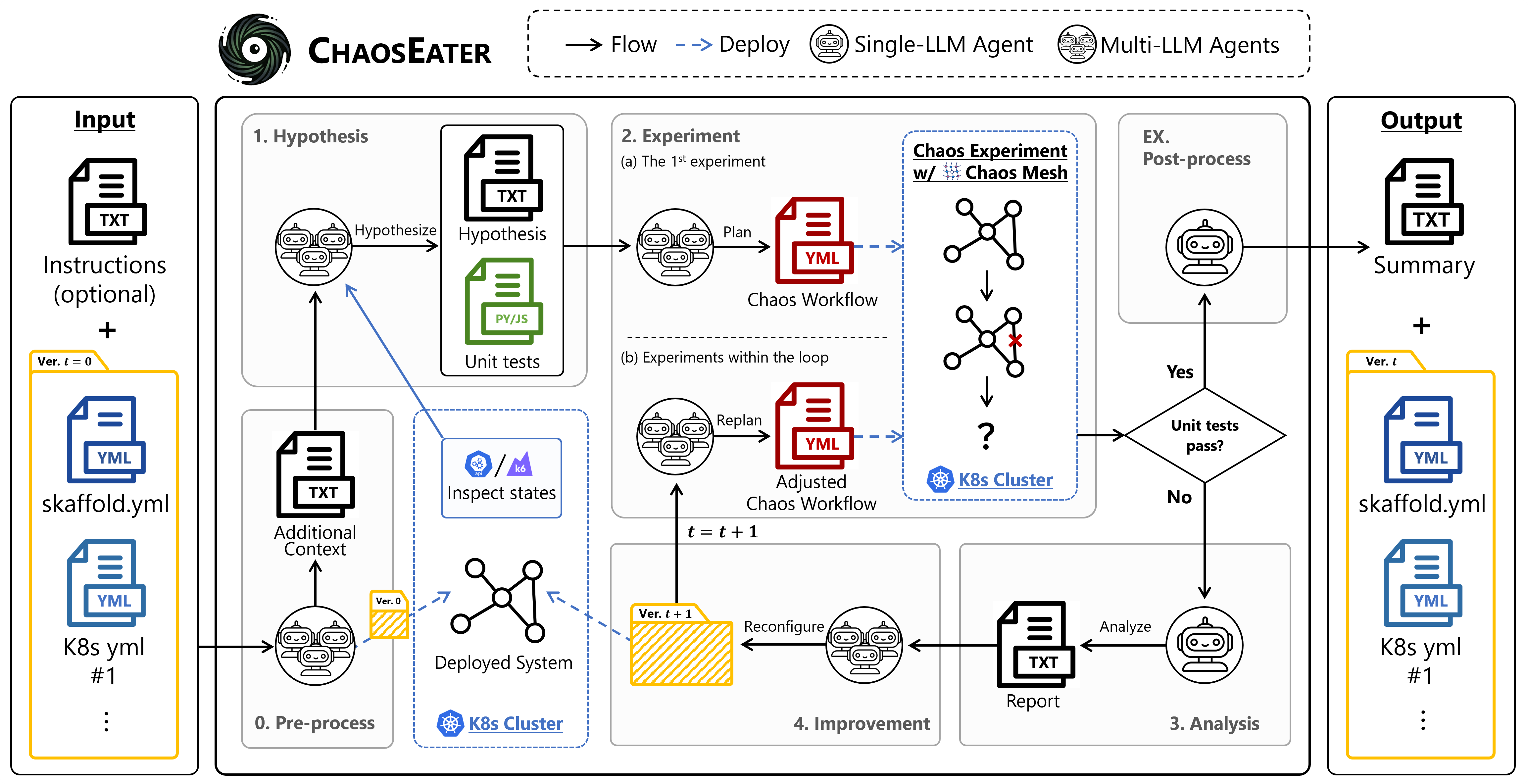}
    \caption{The agentic workflow of ChaosEater. It autonomously completes the systematic CE cycle using LLM agents and existing tools. Note that only the representative inputs and outputs of agents are illustrated here. The two K8s clusters within the workflow refer to the same one.}
    \label{fig:arch}
\end{figure*}

\section{Introduction}
% Background
Modern software applications are built on distributed systems, in which the entire system is composed of multiple component services.
This design, known as microservice architecture \cite{microservice}, enables scalable and continuous deployment while supporting the integration of heterogeneous technologies.
On the other hand, the complex dependencies among microservices can cause faults to propagate, and even minor faults may lead to unpredictable and chaotic behavior across the entire system. 
However, proactively predicting and addressing such complex behavior is challenging.

To address this and improve the resilience of distributed systems, numerous organizations, including Netflix, Amazon, and Microsoft, have recently adopted Chaos Engineering (CE) \cite{ce, ce2}.
Its concept is that \textit{rather than predicting the chaotic behavior, let's observe it directly by artificially injecting faults into the system}. 
Based on the actual observation, we can proactively rebuild a new system that is resilient to the assumed faults.
Systematically, CE cycles through four phases for a target system:
\begin{enumerate}
    \item \textbf{\textit{Hypothesis}}: Define steady states (i.e., normal behavior) of the system and a failure scenario. Then, make a hypothesis that \textit{the steady states of the system are maintained even when the failures occur in the scenario}.
    \item \textbf{\textit{(Chaos) Experiment}}: Inject the relevant faults into the system while monitoring its response behavior.
    \item \textbf{\textit{Analysis}}: Analyze the logged data and check if the hypothesis is satisfied. If so, this CE cycle is finished here. If not, move to the next \textit{improvement} phase.
    \item \textbf{\textit{Improvement}}: Reconfigure the system to satisfy the hypothesis. The reconfigured system is tested again in (2) and (3), i.e., repeat (2) to (4) until the hypothesis is satisfied.
\end{enumerate}

% Problem statement
In recent years, several CE tools \cite{chaosmonkey,aws-fis,chaosmesh,azure-chaos} have advanced the automation of chaos-experiment execution. Moreover, monitoring tools \cite{prometheus,k6,opentelemetry} enable automating metric collection, aggregation, and threshold-based testing during chaos experiments. 
Hence, the \textit{experiment} and \textit{analysis} phases have been mostly automated.
On the other hand, other processes such as making a hypothesis, planning a chaos experiment to test the hypothesis, and reconfiguring the system still remain manual.
These manual processes are labor-intensive and require multi-domain expertise; therefore, automating them is expected to enable anyone to build resilient systems at low cost. 
However, due to their complexity and generative nature, such automation has not yet been achieved by existing algorithmic approaches.
% Since these manual processes are generative tasks that require multi-domain knowledge

% The promise of LLMs for CE
We believe that Large Language Models (LLMs) are the key to overcoming this challenge.
LLMs have recently shown promising capabilities across a wide range of tasks, including general natural language processing \cite{general_llm_survey}, software engineering \cite{llm_code,swebench,swe-agent,devin}, and network operations \cite{llm4nw1,llm4nw2,llm4nw4}.
Considering that CE for software systems corresponds to software engineering, these capabilities sufficiently satisfy the requirements of the manual processes in CE.
Moreover, concurrent with our work, the use of LLMs for automating partial processes in software system management is also being actively explored, including root cause analysis \cite{llm4sce,aiopslab,itbench,one-shot-rca,gpt-4-rca,x-lifecycle,agent4rca}, log parsing \cite{chatgpt-logparse,demonstration-free}, and remediation \cite{genkubesec,auto-remediation,10.1145/3689051.3689056}.
Given these situations, the full automation of CE cycles is now becoming feasible.

% Proposed system
To realize this, we propose ChaosEater, an LLM-based system that automates the entire CE cycle.
It predefines an agentic workflow according to the systematic CE cycle and assigns subdivided CE processes within the workflow to each LLM with a specific role. 
The predefined workflow ensures that the multiple LLMs collaboratively complete CE cycles as intended.
Considering the compatibility between software systems and LLMs, ChaosEater especially targets CE on Kubernetes (K8s) systems \cite{k8s}.
Therefore, the LLMs complete CE cycles through SE tasks, including requirement definition, code generation, testing, and debugging.
In this paper, we specifically present the agentic workflow design, how LLMs perform subdivided CE processes within the workflow, and some techniques for consistently completing reasonable CE cycles.
% , a set of system prompts for creating LLM agents for each CE operation, the interface between LLMs and existing CE tools, and a technique for conducting consistent and transparent testing.
% Experimental results
We also evaluate ChaosEater through case studies on both small- and large-scale K8s systems.
The results demonstrate that it consistently completes reasonable single CE cycles with significantly low time and monetary costs (\$0.2--0.8 and 11--25min). The validity of these CE cycles is also confirmed by two human engineers and three different LLMs.

% Summary
The main contributions of this paper are threefold:
\begin{itemize}
    \item We are the first to implement an LLM-based system that automates the entire CE cycle. This implementation provides evidence for the feasibility of a new direction: the full automation of system resilience improvement.
    \item We release all the resources of ChaosEater.\footnote{Available at https://github.com/ntt-dkiku/chaos-eater.} This release provides a concrete foundation for subsequent works in the new direction.
    \item We evaluate ChaosEater quantitatively in terms of cost and stability, and qualitatively in terms of the validity of its CE processes. The results provide its fine-grained potential, limitations, and future directions.
\end{itemize}

\section{Proposed System: ChaosEater}
\label{sec:chaoseater}

% Overview
In this section, we describe a technical overview of ChaosEater.
% this section describes a technical overview of ChaosEater.
Fig. \ref{fig:arch} shows its simplified agentic workflow.
% input/output
It takes as input instructions for the CE cycle (optional) and a folder containing K8s manifests and a Skaffold configuration file \cite{skaffold}.\footnote{K8s manifests are system configuration files that define the resources (i.e., microservices) that constitute a system, while a Skaffold configuration file defines the deployment process of those resources to a K8s cluster. }
It then conducts a CE cycle for these inputs through six divided phases: \textit{pre-processing}, \textit{hypothesis}, \textit{experiment}, \textit{analysis}, \textit{improvement}, and \textit{post-processing} phases.
Finally, it outputs a summary of the completed CE cycle and a modified folder containing K8s manifests that have been reconfigured to satisfy the hypothesis defined in the \textit{hypothesis} phase, along with their corresponding Skaffold configuration file.

% introducing the subsequent sections
In the following, we describe ChaosEater's workflow design from input to output, breaking it down into the six phases.
Note that, in this paper, we refer to each LLM assigned a specific role (i.e., a subdivided CE process) as an LLM agent, and that the underlying LLMs do not require additional fine-tuning.
See the extended version \cite{chaos-eater} for implementation details, including the detailed agentic workflow, system prompt templates, graphical user interface, and system deployment.

% Validation as code
\begin{figure*}[tb]
  \begin{minipage}{0.54\textwidth}
    \begin{tcblisting}{
        colback=white,
        listing only,
        title={\footnotesize (a) VaC script for K8s API (Python)},
        listing options={
            language=Python,
            basicstyle=\footnotesize\ttfamily,
            breaklines=true,
            numbers=none,
            showstringspaces=false
        },
       boxrule=0.3mm,
       left=0mm,
       right=-3mm,
       top=-2mm,
       bottom=-2mm
    }
def check_podcount(label, expected_count, duration):
  consistent_count = True
  for i in range(duration):
    pods = self.v1.list_namespaced_pod(
      namespace='default',
      label_selector=label)
    pod_count = len(pods.items)
    print(f"current pod count: {pod_count}")
    consistent_count = pod_count == expected_count
    if not consistent_count:
      break
    time.sleep(1)
  assert consistent_count,"Pod count was inconsistent"
...
    \end{tcblisting}
  \end{minipage}\hfill
  \begin{minipage}{0.44\textwidth}
    \begin{tcblisting}{
        colback=white,
        listing only,
        title={\footnotesize (b) VaC script for k6 (Javascript)},
        listing options={
            language=java,
            basicstyle=\footnotesize\ttfamily,
            breaklines=true,
            numbers=none,
            showstringspaces=false,
            escapeinside={(*@}{@*)},
        },
        boxrule=0.3mm,
        left=0mm,
        right=-1mm,
        top=-2mm,
        bottom=-2mm
    }
export const options = {
  vus: 10,
  duration: '10s',
  thresholds: {
    http_req_duration: ['p(95)<500'],
  },
};

export default function () {
  const res = http.get('(*@http://example.com@*)');
  check(res, {'status was 200': (r) => 
    r.status == 200 });
  sleep(1);
}
    \end{tcblisting}
  \end{minipage}
  \caption{Examples of VaC scripts to validate steady states.}
  \label{fig:vac_scripts}
\end{figure*}

% ---------------
% Pre-processing
% ---------------
\subsection{Phase 0: Pre-processing}
Given the user input, ChaosEater first deploys the user's system to the K8s cluster by running the Skaffold configuration file.
Then, each LLM agent sequentially fills in the implicit context of the user's input.
The filled context includes summaries of the K8s manifests, their potential issues for resilience and redundancy, and a possible application, which will serve as auxiliary information in the subsequent phases.
The filtering of harmful prompts in user instructions is also performed here.
% Filtering harmful prompts from user instructions is also performed here.

% -----------
% Hypothesis
% -----------
\subsection{Phase 1: Hypothesis}
The \textit{hypothesis} phase defines the required system resilience for an assumed failure scenario.
% (i.e., requirements definition).
% from a fault tolerance perspective.
Following the principles of CE \cite{ce}, ChaosEater first defines steady states and then defines a failure scenario.

\subsubsection*{\textbf{Steady-state definition}}
% definition of a steady state
Steady states are the expected and normal behavior of a system.
Each steady state is defined by a pair of a state value and a threshold, and a steady state is considered satisfied when the state value meets the threshold.
Therefore, the state values must be measurable outputs of the system, such as the number of active resources, error rates, and response time. 
% workflow overview
Given the pre-processed user input, an LLM agent first defines a measurable state critical to maintaining the system's application.
Another agent then inspects the current value of the state using either K8s API or k6 \cite{k6}.
The inspection is conducted by a Python or JavaScript script written by the agent.
Based on the inspected value, an agent defines the threshold for the state, which, according to the definition of a steady state, must be satisfied under the current normal state.
Finally, an agent adds threshold-based assertions to the inspection script to generate a unit test script that validates whether the steady state is satisfied (Fig. \ref{fig:vac_scripts}).
These processes are repeated to list multiple steady states without duplication until an agent determines that the number of steady states is sufficient.
The unit test script is used for mechanically validating the steady state during the \textit{experiment} phase;
we call this approach of having LLMs judge validity through unit test code \textit{Validation as Code} (VaC), which ensures consistency and transparency of the LLM's validation process.
% The process from defining a state to creating its VaC script is repeated to list multiple steady states without duplication until an agent determines that the number of steady states is sufficient.

\subsubsection*{\textbf{Failure definition}}
Given the pre-processed user inputs and the steady states, an LLM agent proposes a failure scenario that may occur in the system (e.g., a surge in access due to a promotional campaign, cyber attack, etc.), and defines a sequence of faults that simulate the scenario.
ChaosEater employs Chaos Mesh \cite{chaosmesh}, a CE tool that can manage chaos experiments on K8s systems through code; therefore, the faults are selected from those supported by Chaos Mesh.
After drafting faults, another agent refines parameters for each fault, such as the scope of the fault injection, the fault sub-type, and the fault strength.
This stepwise fault detailing helps LLMs specify accurate parameter sets as structured JSON outputs.

At this point, the hypothesis can be reframed as \textit{all VaC scripts pass, even when the defined faults are injected.}

% -----------
% Experiment
% -----------
\subsection{Phase 2: (Chaos) Experiment}
The \textit{experiment} phase plans a chaos experiment to validate the hypothesis and executes it. 

\subsubsection*{\textbf{Experiment planning}}
To enable systematic planning, we divide a chaos experiment into three stages: pre-validation, fault-injection, and post-validation.
In pre-validation, VaC scripts validate steady states under normal conditions.
In fault-injection, faults are injected. If some steady states need to be validated concurrently, their corresponding VaC scripts are also run here.
In post-validation, VaC scripts confirm the recovery of steady states after failures.
Given the pre-processed user inputs and the hypothesis, an LLM agent first determines the duration of each stage. Then, other agents determine the VaC scripts and fault injections to be executed in each stage, along with their execution timing and durations.
Finally, an agent writes a summary of the chaos experiment timeline, which is referenced during the \textit{analysis} phase to identify the causes of the hypothesis rejection.
% plan -> Chaos Mesh workflow manifest 
The chaos experiment plan is then converted to a Chaos Mesh workflow manifest, which enables automated fault injection and hypothesis validation via VaC scripts according to the schedule defined in the manifest.

\subsubsection*{\textbf{Experiment replanning}}
% overview
Resource types and metadata of K8s manifests may be reconfigured during the \textit{improvement} phase.
Therefore, inspection targets in VaC scripts and scopes of fault injections must be updated accordingly between the \textit{improvement} phase and the next experiment execution. 
% inspection target change
Given the original and reconfigured K8s manifests, as well as the previous plan, each LLM agent proposes new inspection targets and new scopes of fault injections.
% workflow manifests
Then, a new ChaosMesh workflow manifest is generated by updating the corresponding parts in the previous one.
% NOTE
Note that this update only makes minor adjustments to reflect the changes in K8s manifests, without altering the original intent of the chaos experiment.

\subsubsection*{\textbf{Experiment execution}}
After the Chaos Mesh workflow manifest is generated, ChaosEater applies it to the K8s cluster. 
Then, the scheduled fault injections and hypothesis validation are automatically executed by Chaos Mesh.
In the meantime, ChaosEater simply waits for the experiment to complete.

% ---------
% Analysis
% ---------
\subsection{Phase 3: Analysis}
% TODO: swipe the report and verification parts
After the chaos experiment is finished, ChaosEater mechanically checks whether the VaC scripts have passed.
If all of them have passed, that means the current system already satisfies the hypothesis.
Therefore, ChaosEater finishes the current CE cycle at this point and moves to the \textit{post-processing} phase.
If at least one has failed, ChaosEater moves to the next \textit{improvement} phase after analyzing the experimental results.
In this analysis, given the K8s manifests, the timeline of the chaos experiments, and the list of failed VaC scripts with their logs, an LLM agent identifies the cause of the test failures and then generates a report containing the causes and countermeasures. 

% ------------
% Improvement
% ------------
\subsection{Phase 4: Improvement}
The \textit{improvement} phase reconfigures the K8s manifests to satisfy the hypothesis.
Given the K8s manifests, the hypothesis, the experiment plan, and the improvement loop history, an LLM agent reconfigures the K8s manifests by replacing, creating, or deleting them, so that all the VaC scripts pass in the chaos experiment.
This agent is instructed to gradually increase redundancy at each step to avoid excessive redundancy.

\subsubsection*{\textbf{Improvement loop}}
After the reconfiguration, ChaosEater applies the reconfigured K8s manifests to the K8s cluster.
They are then validated again through the \textit{experiment} and \textit{analysis} phases.
That is, as in the systematic CE cycle, ChaosEater also repeats the \textit{experiment}, \textit{analysis}, \textit{improvement} phases until the hypothesis is satisfied.
We define this loop as \textit{improvement loop}.
The improvement loop history refers to the history of the experimental results, their analysis reports, and their reconfigurations within this improvement loop, which suppresses the repetition of the same reconfiguration.

\subsection{Extra Phase: Post-processing}
After the CE cycle is completed, ChaosEater finalizes its entire process by summarizing the completed CE cycle.
An LLM agent summarizes the user's input and the four completed phases.
Finally, ChaosEater provides the user with the summary of the completed CE cycle and the folder containing K8s manifests that have been reconfigured to satisfy the hypothesis defined in the \textit{hypothesis} phase, along with their Skaffold configuration file.

\section{Case Study}
% Case study 
CE cycles should not be evaluated solely based on whether appropriate reconfigurations are performed; and it is equally important to evaluate whether each phase leading up to the reconfiguration is meaningful.
Therefore, rather than creating a benchmark that focuses on quantitative metrics, we evaluate ChaosEater both quantitatively and qualitatively through in-depth case studies focusing on two critical cases.

% description of example
% Nginx
The first case, \textsc{Nginx}, is a minimal system consisting of two K8s manifests: Pod.yml and Service.yml.
The former defines a Pod including an Nginx server, and the latter defines a Service routing TCP traffic to the Pod.
To verify whether ChaosEater can improve the system when there are resilience issues, we intentionally configure the resource with a non-resilient setting: we set restartPolicy to Never in Pod.yml.
With this configuration, once the Pod goes down, it will never restart, resulting in extended service outages.
% sockshop
The second case, \textsc{SockShop} \cite{sockshop}, is a practical and relatively large-scale e-commerce system that consists of 29 manifests.
The number of replicas of all the Deployments is originally set to one. However, this setting could lead to downtime of the single replica when it goes down. 
To narrow down this resilience issue, we configure all Deployments to have two replicas except for front-end-dep.yml, which remains at one.
This relatively reduces the resilience of the front-end resource.

For these cases, we validate whether ChaosEater correctly identifies and addresses these resilience issues through a reasonable CE cycle.
To maintain the autonomy of ChaosEater, we input only the instruction to keep each chaos experiment within one minute (for \textsc{SockShop}, access methods are also input).
% LLM settinggs (gpt-4o)
We use gpt-4o-2024-08-06 \cite{openai2024gpt4ocard} as the underlying LLMs.
To improve the reproducibility of this case study, its temperature is set to 0.
We run single CE cycles for each target system five times under the same settings.

\begin{table}[tb]
    % \caption{Time and monetary costs of single CE cycles conducted by \textsc{ChaosEater}. The values for each phase are averaged across runs that did not skip that phase, while the values for overall are averaged across runs that involved system reconfiguration. API costs are calculated from the official OpenAI API pricing table in September 2024. Abbreviations of each phase name are as follows: `All' is the overall process; `Pre' is \textit{pre-processing}; `Hyp.' is \textit{hypothesis}; `Expt.' is \textit{experiment}; `Anlys.' is \textit{analysis}; `Imp.' is \textit{improvement}; `Post' is \textit{post-processing}.}
    \caption{Time and monetary costs of single CE cycles by ChaosEater.}
    % API costs are calculated from the official OpenAI API pricing table in September 2024.}
    \begin{center}
    \scriptsize
    \setlength{\tabcolsep}{4.2px}
    \begin{tabular}{cl ccccccc}
        \toprule
        % &\multicolumn{7}{c}{\textsc{Ngnix}} \\
        % \cmidrule(lr){2-8}
        Target &Metric &All &Pre &Hyp. &Expt. &Anlys. &Imp. &Post\\
        \midrule
        \multirow{4}{*}{\textsc{Nginx}} &Input tokens  &59k  &2.6k  &25k   &13k   &4.4k &5.5k &8.2k\\
        &Output tokens &5.9k &0.5k  &2.5k  &1.7k  &0.6k &0.2k &0.4k\\
        &API cost (\$) &0.21 &0.01  &0.09  &0.05  &0.02 &0.02 &0.02\\
        &Time          &11m  &21s   &2.6m  &4.4m  &50s  &12s  &21s \\
        \midrule
        % &\multicolumn{7}{c}{\textsc{SockShop}}\\
        % \cmidrule(lr){2-8}
        % Metric &All &Pre &Hyp. &Expt. &Anlys. &Imp. &Post\\
        % \midrule
        \multirow{4}{*}{\textsc{SockShop}} &Input tokens  &284k &30k  &150k  &57k  &14k  &15k  &18k\\
        &Output tokens &13k  &5.7k &3.8k  &1.8k &0.7k &0.6k &0.5k\\
        &API cost (\$) &0.84 &0.13 &0.41  &0.16 &0.04 &0.04 &0.05\\
        &Time          &25m  &4.6m &4.3m  &3.3m &36s  &4.3m &21s\\
        \bottomrule
    \end{tabular}
    \label{tab:eval-costs}
    \end{center}
\end{table}
\begin{figure*}[tb] \centering
    \includegraphics[width=\linewidth]{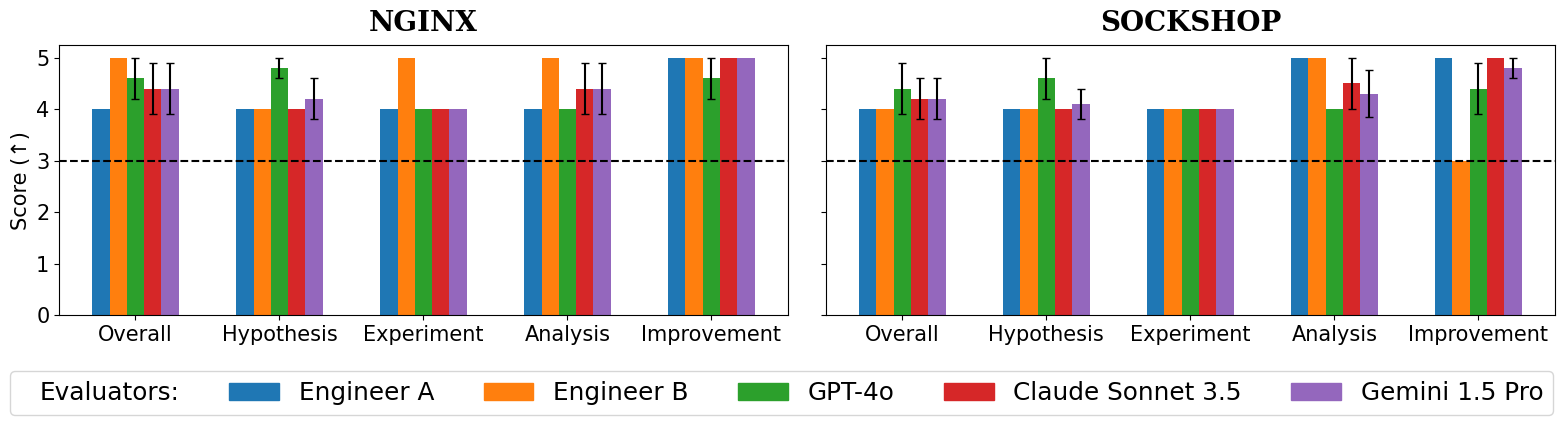}
    \caption{Qualitative evaluation results of CE cycles for each system. A score of 3 or higher is considered a positive rating.}
    \label{fig:qualitative_eval}
\end{figure*}

\subsubsection*{\textbf{Costs and stability}}
Table \ref{tab:eval-costs} shows the time and monetary costs of single CE cycles for each target system.
Although we do not have statistical data on the actual working time and labor costs for the same CE cycles performed by human engineers, the total costs for \textsc{Nginx} (\$0.21 and 11min) are obviously lower than those.
For \textsc{SockShop}, the monetary cost increases by approximately four times (\$0.84), and the time doubled (25min).
However, these values are still intuitively lower than those of human engineers.
Even with the number of resources increasing by more than ten times compared to \textsc{Nginx}, the cost increase remains minimal, demonstrating that ChaosEater maintains low costs even for large-scale systems.
In terms of stability, ChaosEater successfully completes the CE cycle for each system without runtime errors in all five runs. 
It also correctly reconfigures \textsc{Nginx} in all five runs and \textsc{SockShop} in four out of five runs. Even in the non-reconfigured case, we confirm that a valid CE cycle is completed without requiring reconfigurations. 

% -----------
% validation
% -----------
\subsubsection*{\textbf{Qualitative validation}}
To validate CE cycles completed by ChaosEater, we select one of the five runs for each target system and qualitatively evaluate the four phases and the overall process using a five-point scale.
Here, the scale is designed so that a score of 3 or higher is a positive rating.
The evaluators are two external human engineers and three LLMs: GPT-4o, Claude Sonnet 3.5 \cite{Claude3S}, and Gemini Pro 1.5 \cite{gemini}.
The LLM evaluators evaluate each CE cycle five times with a temperature of 0, 
and the final score is calculated as their average.
Fig. \ref{fig:qualitative_eval} shows the results of the qualitative evaluation of the two cycles conducted by the evaluators.
The results show that all evaluators rated every phase above the threshold for a positive rating for both systems, demonstrating that ChaosEater completed reasonable single CE cycles.
In fact, it defined Pod availability as one of the steady states and hypothesized that they are maintained even in scenarios where Pods go down, such as cyberattacks and Black Friday sales. 
Through these experiments, it successfully identified the issues of restartPolicy and the number of replicas, and solved them by replacing Pod with a Deployment resource and increasing the number of replicas.
See the extended version \cite{chaos-eater} for more details of the evaluation settings, the complete outputs, and the full reviews by the evaluators.

\section{Conclusion, Limitations, and Future Work}
CE is labor-intensive and requires multi-domain expertise; therefore, its automation is desirable from the perspectives of cost reduction and accessibility.
Furthermore, as the automatic generation of software applications by LLMs has become widespread recently, such automation is becoming even more important 
for ensuring the resilience of the generated applications.
To address this demand, this paper provided evidence for the feasibility of the full automation of CE, suggesting a future where anyone can build resilient systems at low cost.

On the other hand, the current ChaosEater has several limitations: 
\textbf{1)} It can be deployed only in development environments due to security concerns;
\textbf{2)} It supports only reconfiguration of K8s manifests;
\textbf{3)} Its vulnerability discovery capability is limited within a single CE cycle.

In future work, we will address them in the following directions:
\textbf{1)} It should control the impact range of faults and be safeguarded by a higher-level monitoring system that continuously oversees its operation;
% \textbf{2)} To optimally improve system resilience, it should also reconfigure other types of code, such as HTML, CSS, and JavaScript in the frontend, as well as Python in application code;
\textbf{2)} To optimally improve system resilience, it should also reconfigure frontend code (HTML, CSS, and JavaScript), application code (Python), and lower-layer code (Terraform);
\textbf{3)} It should perform long-horizon vulnerability exploration across multiple CE cycles, where agents iteratively refine hypotheses based on previous cycles and the temporally changing state of the system.

\bibliographystyle{IEEEtran}
\bibliography{main}
\end{document}